# Staying ahead of the curve: progress in British variable star astronomy

**Jeremy Shears**

## Abstract

The BAA Variable Star Section is the world's longest established organisation for the systematic observation of variable stars, having been formed in 1890. Its database contains nearly 3 million measurements going back to 1840 and is an important resource for researchers. The aim of this Presidential Address is to reveal some of the lesser known tales that lie deep within the database. This includes bringing to life stories about some of the people that were involved, especially in the early years, including Joseph Baxendell, Mary Adela Blagg and Arthur Stanley Williams, as well as shedding light on some of the stars that have been observed. Finally we look to the future as the Variable Star Section builds on the legacy of its forebears, ensuring that it shall always stay ahead of the curve.

## Introduction

The BAA Variable Star Section (VSS) is proud to be the world's longest established organisation for the systematic observation of variable stars, having been established in 1890, within a few days of the formation of the BAA itself. From the outset, the primary aim of the Section has always been to encourage the observation of variable stars and to make the resulting data available for analysis. In the early days, little was known about what a star was made of and it was anticipated that the study of the variables would shed light on the internal constitution of the stars. The new astronomy of astrophysics was dawning, supported by its two new observational techniques: spectroscopy and photography. Spectroscopy promised to unlock an understanding of the chemical composition of stars and photography was allowing many more variables to be discovered. The challenge was to enlist a sufficiently large body of astronomers to monitor the stars and derive their light curves. By and large this responsibility was to fall to the vast ranks of amateur astronomers around the world and, in Britain, to the BAA VSS. Other variable star organisations emerged later, notably the AAVSO, which began in 1911.

In the early years of the VSS when there were only a dozen or so stars on the its programme, magnitude estimates were published in the BAA *Journal* with little or no analysis. However, a rapid growth in observations occurred during the first decades of the twentieth century, encouraged by the dynamic section Director, Col Markwick who extended the programme and drew many new observers into the fold. This gave rise to analyses of the results which were regularly presented in the *Journal*. The inevitable result was that within a few years the volume of data was too great for it to be practical to publish all observations in the *Journal*. The solution was to produce a series of *Memoirs*, containing all observations and selected light curves, each carefully drawn by hand. Eventually the sheer volume of results, and the associated publishing costs, rendered this approach unaffordable and the final *Memoir* appeared in microfilm format in 1958, covering the period 1930 to 1934. Few copies of this Memoir were produced and most were deposited at libraries where specialised microfilm readers were available. The consequence was that the information was largely inaccessible to those that might have wanted it.

Fast forward to the 1980's, the advent of the microcomputer meant that observations could be stored electronically and readily made available to anyone that wanted to work with the data. The first viable computerised database was created by Dave McAdam and rolled out in 1991. This was state of the art for the time and allowed impressive light curves to be plotted with ease (Figure 1). Not only were members able to submit their observations to the database, but a mammoth task was initiated of typing in the Section's historical paper records of observations. This was performed by volunteers, who entered the data into their computers at home, and by 1997 the database reached the milestone of 1 million observations. The millionth observation was an estimate of the dwarf nova, SS Cyg made by Mike Gainsford using his 25 cm reflector (Figure 2) on 1996 December 23 at 18.47 UT (1). For the record, the star was at magnitude 9.4 and was just declining from an outburst.

The 2 millionth visual observation was loaded into the database on 2016 March 18. It was an historical observation made by John Toone of Y Lyn (a semi-regular variable of type SRc) at magnitude 7.2 on 2004 December 19. This was part of the ongoing project to fill in the gaps in the observational archive, and was uploaded by VSS secretary Bob Dryden. The graph in Figure 3 shows the cumulative number of observations since 1840. When will we make the 3 millionth observation? And who will be the observer?

Thanks to the valiant efforts of the volunteers, the digitisation of historical visual observations was largely complete by 2013. Further observations are still coming to light, for example in original notebooks or from other archives, and the digitisation of these continues.

With the new technology of CCD photometry becoming more popular in the 1990's, it soon became clear that these data would need to be incorporated into a modified database. The task fell to Andrew Wilson who created the first CCD database in 2004. The first CCD measurement submitted to the database was photometry of supernova 1993J by Martin Mobberley which he extracted from an image he took on 1994 Jan 17 (Figure 4). The image was taken with Martin's 49cm f/4.5 Newtonian to which he had coupled Terry Platt's third ever commercial Starlight Xpress CCD camera.

The latest version of the database, combining both visual and CCD observations, also developed by Andrew Wilson, went online in 2012 and there have been a number of improvements since. Andrew was awarded the Association's Merlin medal for his significant work in developing the database in 2015 (Figure 5).

The VSS database is an important source of data for both professional and amateur variable star researchers. It currently contains about 2.8 million observations (visual, CCD, photographic and DSLR) of some 2900 stars. The most prolific visual observer by far is Gary Poyner, who has more than 281,000 visual observations under his belt. Also in the top 3 are John Toone (>175,000) and Tony Markham (>150,000). David Boyd leads the CCD observations (>270,000) followed by Ian Miller (>210,000) and Roger Pickard (>150,000). Increasingly DSLR photometry is appearing, particularly of eclipsing binaries, where the field is led by James Screech and Des Loughney.

So these are the *facts* relating to the VSS database, which contains a wealth of astronomical data. But to me it is so much more than a collection of digitised data. Within it lie many stories for those willing to look beneath the surface: tales about the observers themselves and about the stars they were so committed to. This Presidential Address will attempt to unlock and reveal some of these stories. It is not an extensive history of the VSS and neither is it intended to be a treatise on the astrophysics of variable stars. These matters are dealt with elsewhere. Instead my aim is to reveal some of the lesser known tales that lie deep within the database. This includes bringing to life stories about some of the people that were involved, especially in the early years, and shedding light on some of the stars that have been observed.

**From the mists of time: Joseph Baxendell**

Although the VSS was formed in 1890, when on October 31 the first Director, John Ellard Gore (1845-1910), presented his proposal for the work of the Section to BAA Council, variable star astronomy in the United Kingdom goes back much further. Sir John Herschel (1792-1871) was the first to promote the systematic observation of variables in the English language in his Treatise of Astronomy, where he noted that: "This branch of astronomy which has been too little followed up, and is precisely that in which amateurs of the science, provided with good eyes, or moderate instruments, might employ their time to excellent advantage" (2). The first observation in the VSS database dates from not long after Herschel's book first appeared. It was made by Joseph Baxendell (1815-1887) on 1840 February 10 and the star was the long period variable (LPV) R Leonis (3). This was an auspicious day as it was also the day when the young Queen Victoria married her cousin, Prince Albert of Saxe-Coburg and Gotha, in the Royal Chapel at St James's Palace.

So who was Joseph Baxendell whose variable star estimates began such an impressive line of observations spanning more than 175 years? Baxendell (Figure 6) was born in Manchester on 1815 April 19, some two months before the Battle of Waterloo. He was educated in Cheetham Hill, Manchester, and at the tender age of 14 was sent to sea by his father who hoped it would invigorate his weakly constitution (4). Clearly parental care was of a different dimension on those days! During a voyage to Valparaiso, whilst off the coast of Central America, he was lucky enough to witness the remarkable Leonid meteor storm of 1833 November 13 when the night sky was briefly filled with meteors. Whilst on board ship he taught himself the essentials of mathematics (5).

In 1835 Baxendell left his sea-faring career behind and returned to Manchester where he assisted his father who was a land steward and Joseph subsequently established his own estate agency business. With his friend Robert Worthington, the young Baxendell set up an observatory with a 13-inch (33 cm) reflector (the speculum of which Baxendell cast, ground and polished) and a 5-inch (12.5 cm) refractor at Crumpsall Old Hall near Cheetham Hill, Manchester. Worthington was partially sighted as a result of an accident to his right eye and, although still able to observe, he was happy to let Baxendell have the run of the observatory.

Baxendell mainly focused his observational work on variable stars, discovering some 16 new variables, including λ Tauri in 1848 (6), which was only the third eclipsing binary to be discovered, after Algol and β Lyrae. This was quite an observational feat considering the amplitude of λ Tau is only about 0.5 mag (7). These discoveries brought him to the attention of the astronomical community at home and abroad. The VSS database contains observations 7,586 by Baxendell (3). Whilst much of his work was focused on variable stars, he also observed other objects, including Jupiter on which he followed an unusual oblique streak in 1860 and which was reported by several other observers (8).

In 1858 Baxendell joined the Manchester Literary and Philosophical Society which was established in 1781. He held several senior positions in the "Lit & Phil", soon joining its council, and in 1861 he became joint secretary as well as editor of the *Proceedings*, holding the latter post until his death. This was a Golden Age for Manchester science and technology (9) and Baxendell's contemporaries included James Prescott Joule (1818-1889) of Salford, discoverer of the mechanical equivalent of heat, and Balfour Stewart (1828 –1887), Professor of physics at Owens College (now part of the University of Manchester) and previously of the Kew Observatory. Balfour Stewart became a close friend of Baxendell and it was he who wrote Baxendell's obituary in *Nature* (5).

Baxendell was appointed Astronomer of the Corporation of Manchester in 1859. Whilst this might be perceived as a luxury for a city with a reputation for cloud, one of his main duties was to maintain a time service for the city, which was especially important as the railways developed, connecting Manchester to the rest of the country and beyond. At the time Manchester and its environs, or "Cottonopolis", was not only an industrial and manufacturing powerhouse of the North, but also of the world.

Baxendell collaborated extensively with two other prominent variable star observers of the time, Norman Pogson (1829-1891) and George Knott (1835-1894). Pogson was employed at a number of observatories during his career, including George

Bishop's observatory in Regents Park, the Radcliffe Observatory at Oxford and John Lee's Hartwell Observatory. In 1860 he was appointed government astronomer at the Madras Observatory, which position he held for 30 years. There was also a family connexion as Baxendell married Pogson's sister, Mary Anne, in 1865.

Baxendell and Knott attempted to set up the *Association for the Systematic Observation of Variable Stars* in 1863, but it never took off. A comprehensive review of the development of British variable star associations, leading to the formation of the BAA VSS is presented in an excellent paper by John Toone in an earlier edition of the *Journal* (10)

In later years, Baxendell resided at the seaside resort of Southport where, in 1871, he was appointed Superintendent of the Meteorological Observatory in Hesketh Park. This was set up by John Fernley (1796-1873), one of Southport's great benefactors, who also donated the meteorological apparatus. Baxendell erected a private observatory at his residence in Liverpool Road, Birkdale, Southport, in 1877 from where he continued his observations of variable stars with a fine 6-inch (15 cm) f/14.6 equatorial refractor by Cooke & Sons (Figures 7a and b). The observatory, the equatorial, a micrometer, a portable transit instrument and a sidereal chronometer were the property of Sir Thomas S. Bazley (1797-1885) of Hatherop Castle, Fairford, Gloucestershire, who loaned them to Baxendell, even paying for them to be removed from Fairford and re-installed at Birkdale. Bazley was a Manchester industrialist and mill owner, MP for Manchester between 1858-1880 as well as a member of the Lit & Phil, so would have known Baxendell well (11). He was a generous benefactor in Manchester.

Baxendell's son, Joseph Baxendell Jr., who eventually took over from his father as Superintendent of the Meteorological Observatory and for a period assisted his father in making observations at Birkdale, later donated the telescope to the Education Department of Southport Corporation. It was set up in the Fernley Observatory in Hesketh Park at Southport where it stands today (Figure 8).

In 1884 Baxendell was elected Fellow of the Royal Society. He didn't live to see the formation of the BAA, but there is no doubt that had he done so he would have been an enthusiastic member, having been a member of the Liverpool Astronomical Society.

**Light curve analysis: Mary Adela Blagg**

It is the hope of all variable star observers that their magnitude estimates, gleaned through months and years of hard work, will be combined with the results of other observers and analysed to reveal new insights into the stars under study. In fact, for many of us this is one of our prime motivations to go out and commune with the variables night after night, no matter how cold or damp it might be. During the first few decades of the VSS, little analysis was performed beyond plotting light curves and estimating the times of maximum of stars, especially the LPVs where the aim was to characterise the cycle length between successive maxima and determine whether this, and the brightness at maximum and minimum, changes over time. In many cases these parameters were determined by eye by careful inspection on the light curve (Figure 9). One person to turn their hand to mathematical analysis of light curves was Mary Adela Blagg (1858-1944). Never a systematic observer of variables herself, she turned her evident intellect to a more thorough analysis of their variation.

Mary Blagg (Figure 10) and her important contributions to both variable star astronomy and selenography deserve to be better known (12). Even during her life very few people met her: she rarely attended meetings and, because of a modest and retiring disposition, seldom ventured beyond her Staffordshire home. However, she was among the first group of women to be admitted as Fellows of the RAS on 1916 January 14. This was an incredibly important, yet long awaited, step for the RAS. Whilst some women had been allowed to attend meetings, Fellowship was not open to them. Attempts were made over the years to change the rules to allow the admission of women. Eventually there was an overwhelming vote in favour at its 1915 AGM and the RAS organised the approval by the Privy Council of a Supplemental Charter, which paved the way for women finally to be admitted to Fellowship.

The other women admitted on the same day as Mary Blagg were Miss Ella K Church (13), Miss Irene Warner, Miss Alice Grace Cook and Mrs Fiametta Wilson. It is important to note that all were BAA members, the BAA having a much more enlightened attitude and had admitted women since it began in 1890. Grace Cook was an enthusiastic variable star observer, making an early independent discovery of Nova Aquilae 1918 and it was Fiammetta Wilson who took future VSS Director Félix de Roy under her wing following his escape from Belgium to London at the beginning of the First World War, lending him a telescope so he could continue his observations during his stay in Britain. Fiammetta Wilson and Grace Cook took over the running of the BAA Meteor Section during World War 1 as the Director, Rev. Martin Davidson, was away serving as an Army chaplain.

Mary was born on 1858 May 17 at Cheadle in Staffordshire, the eldest daughter of well-to-do solicitor Charles John Blagg (1833-1915). She was educated at home and at a private boarding school in London. She did not attend college or follow a profession, but became involved voluntary work, including in the local church and the Cheadle Girls Friendly Society. After the death of her mother, Frances, in 1896 Mary assumed responsibility for the household at Greenhill, the large family home (Figure 11). During the First World War she looked after Belgian refugee children that were evacuated to Cheadle, a role which gave her much pleasure (Figure 12).

Early on it became clear that Mary had a natural aptitude for mathematics and expanded her knowledge in the subject by borrowing her brothers' school books. She became especially interested in the mathematics of harmonic analysis which was later to stand her in good stead in her variable star analysis. However, it was only when she approached middle age that Mary developed an interest in astronomy through attending a course of University Extension lectures given at Cheadle by Joseph Alfred Hardcastle (1868-1917; Figure 13).

J.A. Hardcastle, a grandson of Sir John Herschel, became a BAA member in 1902 and later served as Secretary. Here he met S.A. Saunder (1852−1912), who served as BAA President 1902- 1904, and was a mathematics teacher at Wellington College. Saunder was engaged in a project to measure accurate positions of lunar craters from photographs taken at the Paris Observatory and he invited Hardcastle to join him in this enterprise. However, they soon became frustrated by the lack of a standardised nomenclature for the lunar craters that were already known; confusingly, many were known by several different names. Professor H.H. Turner (1861-1930) of Oxford University realised that standardisation was required and gained support from the Royal Society and the International Association of Academies at its meeting in 1907 to produce a standard list of formations for use by selenographers.

The work was detailed and tedious and it soon became clear to Saunder that he needed additional help if he was to achieve anything of lasting value. Hardcastle recognised Mary Blagg's keen eye for detail, having communicated to the BAA *Journal* a short paper she had written on *The Scintillation of Stars* based on a year's worth of careful observations she had made from Cheadle during 1905 (14). He therefore suggested that she might like to become involved in the project. Saunder and Blagg worked carefully on this painstaking venture for several years and in 1913 they published the definitive *Collated list of lunar formations* (15) (Figure 14). Over the years, Blagg produced further lists and began to cooperate with European selenographers on an expanded publication which came out under the auspices of the International Astronomical Union (IAU) in 1935, *Named Lunar Formations* (16).

Blagg's work on lunar nomenclature brought her to the attention of H.H. Turner. In the same year as her *Collated list* was published, Turner communicated a paper on her behalf to the *Monthly Notices* on a *Substitute for Bode's Law* (17). The core of the paper was *Blagg's Formula*, as it became known, which was a generalised form of Bode's law, an empirical rule giving the approximate distances of planets from the Sun, but which can equally well applied to planetary satellite systems.

In 1907 Joseph Baxendell Jnr presented his father's original variable star observations going back to 1836 to H.H. Turner (18). Turner immediately recognised that, because the data were so voluminous, he needed help with their analysis: (19)

"It is scarcely possible to hand it to anyone who is not familiar with variable star records….. I take this opportunity of saying that I should be very glad of skilled volunteer assistance, at any rate in dealing with the copied ledgers for different stars, and perhaps with these early records also. If any variable star observer has leisure for work of the kind and would communicate with me, I should gratefully accept

4assistance in making this mass of valuable material ready for publication as soon as possible. Unaided, my work at it must necessarily be slow".

Mary Blagg immediately volunteered to assist Turner in the project. Initially the two worked together and some ten papers in the *Monthly Notices* resulted. This was tedious work as it involved tabulating Baxendell's observations, determining which comparison stars he had used and what their brightness was from the Harvard College photometry, then finally reducing the light estimate to obtain the magnitude of the star at the time it was observed. Only then could a mathematical analysis on the resulting data begin. The first paper was on R Arietis and was published under Turner's name in 1912, but in the text he freely acknowledged that Blagg had done most of the work presented. He said "I may perhaps be allowed to express my regret that Miss Blagg does not wish her name to appear as joint editor. The great part of the work is due to her" (20).

Clearly enthused by variable stars, Blagg went on independently to analyse and publish observations from VSS members, notably a series of studies on the LPVs RT Cyg, V Cas and U Per, using observations made by VSS Secretary Arthur Neville Brown (1864−1934), producing new elements for the periods of these stars. Brown's light curve of RT Cyg is shown in Figure 9.

A star that Blagg took particular interest in was the eclipsing binary β Lyrae, which had been discovered by the pioneering variable star observer, John Goodricke (1764-1786) of York in 1784. During the 1840s and 50s, Argelander noticed that the period between eclipses derived by Goodricke was changing and he gave ephemerides with second and third order terms. Blagg undertook a thorough analysis of the times of minimum initially with observations made by Baxendell Sr. between 1840-77 (21). She then extended her analysis with data obtained by VSS members between 1906-1920 (22) and in a third paper (23) she included additional VSS data (1898-1905) as well as magnitude estimates from other observers. This resulted in the most comprehensive analysis of β Lyr in the 136 years since its discovery and which confirmed that the period was indeed increasing in a consistent manner.

An analysis reported in 2005 of minimum times of β Lyr going back to Goodricke's measurements show that the orbital period is increasing at a rate of about 19 seconds per year. The O-C diagram (24) of eclipse timings shown in Figure 15 exhibits a parabolic shape, indicating a constant rate of period change. We now understand that this is because of continuous mass transfer between the two stars of about $2 \times 10^{-5}$ solar masses per year. β Lyr is a semi-detached binary system made up of a B-type main sequence star (the "donor") which is losing mass to a "gainer" B-type star of around 13 solar masses. It is thought that the gainer is completely embedded in a thick accretion disk, which contributes about 20% of the light of the system, with a bipolar jet-like structure perpendicular to this disk, which creates a light-scattering halo above its poles. The system has recently been imaged using the CHARA interferometer array on Mount Wilson which resolved the component stars (25).

In addition to the regular eclipses, β Lyr shows smaller and slower variations in brightness. These are thought to be caused by changes in the accretion disc and are accompanied by a variation in the profile and strength of spectral lines, particularly the emission lines. The variations are not completely regular, but have been a period of about 282 days (26)

Blagg's work on selenography and variable stars brought her international recognition and she was persuaded to leave Cheadle to attend the IAU meetings in Cambridge (1925) and Leiden (1928). At the Cambridge meeting she was invited to join IAU Commission 27 on Variable Stars where she encountered many famous names in variable star astronomy and the world of astrophysics, including Arthur Eddington, Enjar Hertzsprung, Joel Stebbins and William Tyler Olcott (who founded the AAVSO in 1911). No less famous were the astronomers she met in the IAU Lunar Commission meetings of which she was also a member. These include Guillaume Bigourdan, Karl H. Müller and W.H. Pickering. It must have been quite an adventure for this quiet and unassuming woman from Staffordshire.

In later years, Mary moved from Greenhill into a cottage in Cheadle, "High Bank", which was built for her and her unmarried sister Dorothea. Mary suffered from heart problems in her final years and passed away in 1944 at the age of 88. In her obituary, P.M. Ryves noted that "she was far more than a mere amanuensis or compiler of facts….in all her work she displayed not only skill and judgement, but also originality and courage" (12). Clearly Blagg had all the attributes that could have enabled her to become a professional astronomer had circumstances been different.

Given her work in selenography, it is also fitting that following her death a lunar crater was named after her. This is a 5 km impact crater located on the Sinus Medii not far from the Triesnecker crater and the rille system (Figure 16a and b). In 2001, as part of the Millennium celebrations, her home town of Cheadle honoured their brilliant daughter by erecting an armillary sphere in her memory. This was designed and constructed by BAA honorary member, and Cheadle blacksmith and engineer, Mr James Plant (Figure 17). It is fitting that we recognise the work of Mary Adela Blagg in this centenary year of her Fellowship of the RAS.

**Variable star discoveries and Arthur Stanley Williams**

The closing years of the nineteenth century and the beginning of the twentieth saw a surge in the number of variable star discoveries. This was largely driven by the emerging photographic surveys, notably the one conducted at Harvard College Observatory which began in 1885, but also by visual observers systematically searching for them by visual means.

An early adopter in amateur circles of photographic research on variable stars was Arthur Stanley Williams (1861-1938). Stanley Williams (Figure 18), an original member of the BAA, is perhaps best remembered for his planetary observations, with a particular interest in Jupiter (Figure 19), although he also made extensive observations of Mars and Saturn. His variable star research has been somewhat overlooked, even though his award of the RAS's Jackson-Gwilt medal in 1923 cites his contribution to both variable star and planetary astronomy.

Stanley Williams was a solicitor by profession, although his retiring disposition led him to abandon his legal career to live most of his life as a recluse. His great passion was yachting and most of his cruises were single-handed. In 1920 he won the Royal Cruising Club Challenge Cup - the highest award in the world of cruising - for a single-handed cruise of 1254 miles (2018 km) from Falmouth to Vigo in north-western Spain and back. For many years he lived at St Mawes in Cornwall on a barge named *The Queen*. He cruised the Cornish coast in the summer, but during the winter months he beached the vessel at a spot where he had placed his observatory on the shore nearby.

Williams' visual variable star observations began during a visit to Australia in the (northern) winter of 1885-86 which he later published as *A Catalogue of the Magnitudes of 1081 Stars Lying Between -30° and the South Pole* (27). In the process he discovered several new stars the first being the eclipsing binary, V Puppis. His first attempts at discovering new variables by photography began in 1883. In his notebook he writes: (28)

"The importance and great future of photography in variable star work first occurred to the writer in the latter part of the year 1882, when he heard of the photographs of the great September comet of 1882 which had been obtained by Gill at the Cape Observatory. They were said to show stars down to the ninth magnitude, and they had been made with a portrait lens attached to the tube of one of the equatorials, if rightly remembered. At once it seemed that we had here a great and hitherto unsuspected agent for the observation and detection of variable stars, particularly for the detection of variable stars of small range of variation…"

However, Williams' initial attempts at photography were thwarted by the slow plates available to him at the time. He resumed the project in 1899, by which time faster plates were obtainable, using a 4.4-inch (11 cm) portrait lens by Grubb mounted on a Horne & Thornthwaite equatorial with clock drive. Remarkably he set up the camera in a bedroom in his house in Hove that had a large window, the top half of which he opened whilst making the exposures. The narrow strip of visible sky afforded exposures of up to 1 hour.

Williams discovered around fifty new variable stars in the first four years of his re-launched photographic project. Once he suspected a star of variability on his plates he followed up with visual observations made with his 6½-inch (17 cm) Calver reflector – the same telescope he used for most of his planetary observations.

One of the most important stars that Williams discovered was RX And, which he announced to the world in *Astronomische Nachrichten* in 1905 (29). This is now known as a dwarf nova of the small grouping known as Z Cam systems which undergo "standstills" for extended periods of time between outburst and quiescence (Figure 20). There is another link to the BAA VSS as the Z Cam stars were first proposed as a separate class of object by another pioneering VSS member and Director, Félix de Roy of Belgium. RX And is still monitored by the VSS and other groups around the world (coincidentally, just before this Presidential Address was presented, Professor Christian Knigge of Southampton University organised a campaign amongst amateurs to monitor RX And in support of his observations of the system with the Chandra X-ray satellite during an outburst in 2016 October).

One reason why RX And has attracted much professional attention as during the century since its discovery its behaviour appears to have changed. As well as outbursts and standstills, it occasionally enters low states where the star stops outbursting entirely and fades to about a magnitude below the normal minimum (Figure 21).  Such behaviour - flipping between a "high state" and a "low state" - is similar to that of another family of cataclysmic variable, the VY Scl stars. In fact it has been suggested that RX And might be a transitional object between Z Cam and VY Scl systems (30). As a general rule in astrophysics, transitional examples offer the possibility of revealing important new insights, in this case into the evolution of dwarf novae.

**An uninterrupted visual record: Mr. Gore's nova in Orion**

In addition to observations of R Leo, whose recorded observations, as we have seen, go back to 1840, the VSS database contains long runs of observations spanning well over a century of many other variables, especially the LPV's that formed the core of the programme from the outset.  One example of such a star is U Orionis, which was discovered on the evening of 1885 December 13 by John Ellard Gore (Figure 22). At the time Gore was Director of the Variable Star Section of the Liverpool Astronomical Society, although 5 years later he would become the first Director of the BAA VSS. Observing from his home town of Dublin, Gore detected what was initially taken to be a 6th magnitude Nova in Orion (31):

"I first saw this star with a binocular field-glass.....at 9.20 P.M., Dublin mean time. My attention was attracted to it by its very reddish colour and its absence from Harding's charts"

He informed Ralph Copeland (1837-1905) at Dun Echt Observatory, Aberdeen, of his discovery on December 16. Copeland confirmed it the same evening and sent a telegram of the news to Harvard College Observatory (HCO), USA, as well as distributing a circular alert to other astronomers around the world. HCO picked up the object later the same night at magnitude 6.0. Spectroscopy at HCO and the Royal Observatory Greenwich (32) showed that it was not a nova after all, but an LPV. The star is now known as U Ori and is notable as the first LPV to be identified via a photograph of its spectrum. "Gore's Nova" and "Nova Ori 1885" as still listed in the SIMBAD Astronomical Database.

Gore's examination of observations during the first 5 years since his discovery yielded a period of 373.5 days (33). An analysis by the present author of the 14500 observations in the VSS database between 1885 and 2016 yields a period of 370.5 days and a typical range of magnitude 6.5 to 12.0. U Ori was the first LPV to be shown to be asymmetrical in shape via infra-red observations made during a lunar eclipse (34).

A light curve of U Ori between 1900 and 1919 is shown in Figure 23. However the light curve spanning more than a century shown in Figure 1 reveals a number of points, not so much about the variability of the star, but about the ebb and flow of variable star astronomy in the BAA and even how sociological and geopolitical events can intervene.

First, it can be seen that there are relatively few observations in the 1890's, but this improved during the first decade or two of the twentieth century as new members joined the ranks of the VSS. There appear to be two golden periods: one in the 1920's-30's and then again in the immediate post Apollo era. On the other hand, there was a dramatic decline in the number of observations coinciding with the start of World War Two. In spite of this, some observers continued to observe even when on active service, with estimates received from a number of unusual places, such as El Alamein and Mersa Matruh during the Western Desert campaign. Following the end of hostilities, there was an increase in observational activity in 1946. Sadly, the renewed activity was not sustained. The VSS Director, W.M. Lindley, who had once been one of the most active members of the section was away on military service for much of the war and when it was over he never managed to recover his previous enthusiasm for driving the affairs of the section forward. Moreover, few Section reports were written, which also gave the impression of a rather moribund Section. The situation gradually improved following the appointment of R.G. Andrews as the new Director in 1958.

**The R Coronae Borealis stars**

What is the most observed star in the VSS database? In view of the attention given by the VSS to LPVs, especially in the early years, it might be expected that a member of this class might be the one with the largest number of observations in the database. Indeed there are many LPVs with well over one hundred years of continuous observations. However, because their brightness generally changes rather slowly, best practise dictates that they should be observed no more frequently than every few days or perhaps once a week. By contrast, other types of variable can show nightly or intra-nightly variations, and therefore should be observed every clear night or even more frequently. These include, for example, the dwarf novae and the R Coronae Borealis stars and it is R CrB itself that has the honour of being the star with the greatest number of observations. The total stands at more than 53000 visual estimates (actually, two dwarf novae are the second and third most observed: SS Cygni, with more than 47000, and SS Aurigae with more than 31000).

The light curves of R CrB stars show dramatic and unpredictable fades of several magnitudes that occur within a few weeks. Over succeeding months, they gradually recover their original brightness. R CrB stars are highly evolved and their surfaces are unusually poor in hydrogen, but rich in heavier elements including carbon and nitrogen. The fades are now known to be caused by condensation of carbon, which is essentially soot, making the star fade in visible light, while measurements in infrared light exhibit no real luminosity decrease.

Whilst R CrB might be the archetype of the family, having been discovered by Edward Pigott (1753-1825) of York in 1795, it is another member of the family that has a particular connexion with the BAA VSS, RY Sagittarii.

RY Sgr was discovered by E.E. Marwick whilst serving in Gibraltar as an officer in the army Ordnance Department. In a systematic patrol for new variable stars and novae, he noted RY Sgr in his binoculars as 7th magnitude in July 1893. However, when he next returned to the field in September, he was no longer able to see it (thus putting it fainter than mag 9) (35). In the HCO Circular no. 7 dated 5 June 1896, E.C. Pickering described the events surrounding the discovery, which was independently made by Markwick and Williamina Fleming (36) (1857-1911) at HCO: (37)

"[RY Sgr] is a very remarkable object. It was one of a list of 42 stars suspected of variability, sent here for examination, by Col. Markwick. A report was sent to him that an examination of several photographs failed to show any sign of variability. A few days later an object having a peculiar spectrum was discovered by Mrs. Fleming. All the plates of the region were examined and its variability established. It was about to be published in [HCO] Circular No. 6, when it was found to be identical with the star of Colonel Markwick. It was accordingly reported to him for announcement, but he has kindly authorised its publication here".

It was with some degree of satisfaction that Markwick, quite rightly, realised that his modest search with a small "regulation pattern" binocular (price two guineas) had yielded a result comparable to that achieved by the well-resourced HCO with its photographic patrol and supporting team of researchers.

Figure 24 displays a light curve based on more recent VSS observations of RY Sgr and which shows the fade that took place during 1999.

Careful analysis of the light curve of RY Sgr between 1920-1932 revealed a secondary Cepheid-like variation during maximum light of about 39 days and with amplitude of about half a magnitude (38). Recent studies suggest that the pulsation period of RY Sgr is decreasing (39). It has been suggested that this change might be the result of rapid evolution of the post-giant phase of the helium-enriched star; alternatively it may be due to significant mass loss. It is clear that this star still has much to tell astrophysicists. As ever, further observations by the amateur community are required to find out what will happen next to the star!

We cannot mention Markwick without highlighting two innovations he introduced whilst VSS Director and which helped the section to flourish. The first innovation was a concept which is now commonplace in the BAA: the idea of a section meeting where members could get together to discuss section matters and observing. Thus the first VSS meeting was held on 1906 December 10 at a London hotel, attended by 8 members. Such was the apparent success of this meeting that he suggested other sections take up the idea, remarking that one objective of such meetings was "to infuse as much life into our work as possible" (40). His other idea, also formulated with the aim of fostering a spirit of cooperation amongst VSS members, was to distribute section circulars; at least 79 circulars were issued whilst he was Director.

**The Halloween star**

People have asked me what is the most unusual variable star in the VSS database. Indubitably, what constitutes "unusual" is in the eye of the beholder. Each variable is of course unique and there are many categories of variables to choose from; their stories can be gleaned by a careful interrogation of the VSS database. The most exciting aspect of observing variable stars is when the star throws away the script and extemporises. It is for this reason that personally I find the dwarf novae, novae and supernovae exciting because of their sudden and unpredictable outbursts, their exquisite light curves and the fundamental science their observations can unlock. The same can be said for the active galactic nuclei (AGN) where the central region of the galaxy is observed to outshine all the billions of stars in the galaxy itself. For example the AGN known at BL Lac, which was originally classified as a variable star, but we know it is a quasar with a strong jet pointed right at us. Amateur observations of BL Lac and countless other variables have helped in understanding the nature of the objects themselves.

For me the most unusual object in the database, and one that I have observed, brings the story of the VSS right up to date. And curiously it concerns a star whose inherent brightness doesn't change at all! The object first came to public attention with a missive from the IAU's Central Bureau for Astronomical Telegrams on 2006 October 31 - hence the moniker of "the Halloween Star". This announced that the Japanese nova hunter, Akihiko Tago, had reported that a star in Cassiopeia, GSC 3656-1328, had brightened from mag 11.8 to mag 7.5. It was amateur astronomers who first confirmed Tago's observation and their observations began to appear online on the various variable star observing groups within a few hours. I remember dashing home from work on October 31 to open up the observatory, noting that the first trick-or-treaters were venturing outside. It was not properly dark when I started to download CCD images of the field of the new object, provisionally referred to as Var Cas 06. In spite of the noisy images, to my amazement a bright object was staring back at me!

Time-series photometry by amateurs belonging to the VSS, the AAVSO and the informal grouping known as the Center for Backyard Astrophysics began immediately. Anticipating that this would be the outburst of a dwarf nova, or perhaps even a nova, as near continuous photometry as possible from multiple longitudes was needed. However, this simply revealed that the star was gradually fading. There were none of the humps and bumps in the light curve that might be typical of a cataclysmic variable. Moreover, when spectroscopy with large telescopes and X-ray observations from the Swift satellite became available they also revealed the object lacked the characteristics of a cataclysmic variable: there was no change in its spectrum during the outburst and no discernible X-ray signature. A search of archival plates from 1964 to 1994 showed no variability in the star during the 30-year interval (41).

So what was happening? The object was unlike any other known type of intrinsic variable star and it was therefore quite a mystery. Well, the first suggestion that this could be a rare gravitational microlensing event was made a good friend of the VSS, Dr. Chris Lloyd, on November 4. A search for observations made in the lead up to the peak brightness on October 31 revealed several pre-discovery observations, some by amateur astronomers and some by the All Sky Automated Survey (ASAS). The resulting symmetrical light curve (Figure 25) was entirely consistent with a microlens event and it was the nearest one ever recorded.

Gravitational microlensing occurs when the gravitational field of a star acts like a lens, magnifying the light of a distant background star when the two stars are almost exactly aligned along the line of sight. Albert Einstein had introduced the concept of gravitational microlensing of one star by another closely aligned along its line of site in 1936, although he concluded "there is no great chance of observing this phenomenon" (42). Searches for microlensing didn't get underway until the 1990's and several thousand were discovered toward several lines of sight, including the Large and Small Magellanic Clouds. More than a thousand such events have been observed over the past ten years, but these are typically of very faint stars. Gravitational lensing is widely used to detect exoplanets. If the foreground lensing star has a planet, then that planet's own gravitational field can make a detectable contribution to the lensing effect. Extragalactic gravitational lenses involving galaxies and quasars are well known, such as Einstein Cross and Einstein Ring.

What caused this particularly gravitation lens event on Halloween 2006? The lens star is a faint low-mass star or brown dwarf, about one-sixth of the mass of the Sun and with a relatively high proper motion, located at a distance of ~130 pc. It is thought to have passed in front of GSC 3656-1328 which is much further away at a distance of ~1 kpc. The chances of such an exquisite alignment are very small indeed and it has been suggested that a comparable microlensing event of a 12$^{th}$ magnitude star in the Tycho-2 catalogue of 2 million stars might occur every 6 to 12 years (41). It will be interesting to see whether the large scale sky surveys that are coming on line, such as Gaia, will detect further such events.

It is worth noting that even in the 21$^{st}$ century, amateur observers were key to the discovery, follow-up and interpretation of this rare event.

**Looking to the future**

It is often said that astronomy is one of the few remaining sciences where amateurs can still contribute to research. The study of variable stars by amateur astronomers is often invoked as an example. The BAA VSS has played an important role in variable star astronomy, contributing important observations which have helped push back the frontiers for the last 126 years, as I hope this article has shown. It is a field of astronomy that is open to all amateur astronomers, whether they observe with their naked eye or with a sophistic telescope and CCD camera. Combining observations made by amateur astronomers from around the world – for this is an international endeavour - can certainly yield new astrophysical insights, some of which have been described here.

With the advent of large synoptic surveys, such as Gaia, Pan-STARRS and LSST, many more variables will be discovered and once again their follow-up with fall to amateur observers. It is essential that as we look to the future we continue to build on the legacy of our forebears, thus ensuring that we shall always stay ahead of the curve!

**Acknowledgements**

Richard Masefield, whose great-grandmother's eldest sister was Mary Blagg, and James Bradbeer Kindly provided information about Mary Blagg's life and activities. Dr Richard McKim generously gave information about Stanley Williams and allowed me to reproduce Williams's photograph. James Plant kindly hosted a visit to Cheadle during which I was able to view his wonderful memorial to Mary Blagg.

In addition to extensive use of the VSS Database, this research made use of the NASA/Smithsonian Astrophysics Data System, the AAVSO Variable Star Index, the AAVSO International Database, and the SIMBAD Astronomical Database operated through the Centre de Données Astronomiques (Strasbourg, France).

Last, but by no means least, I thank the observers, past and present, who have contributed observations to the VSS database.

Address: "Pemberton", School Lane, Bunbury, Tarporley, Cheshire, CW6 9NR, UK
[bunburyobservatory@hotmail.com]

43. *This photograph was also published as a plate in JBAA, 49, facing page 359 (1939), as part of Williams' obituary.*

44. *JBAA, 1, 174 (1891).*

45. *Howarth I.D., JBAA, 87, 395 (1977).*

46. *Observers: A. B. Burbeck, A. N. Brown, C. F. Butterworth, C L Brook, E. E. Markwick, E. J. How, F. de Roy, F. R. Cripps, G. B. Lacchini, G. W. Middleton, H. Corder, H. L. Dilks, H. Thomson, J. A. Greenwood, J. H. Bridger, J. Kelly, J. M. Offord,. J. W. L. Child, M. A. Orr, M. E. J. Gheury, N. V. Ginori, O. A. Lebeau, P. M. Ryves, R. L. Waterfield, T. W. Backhouse, W. M. Worsell., W. T. Gayfer.*

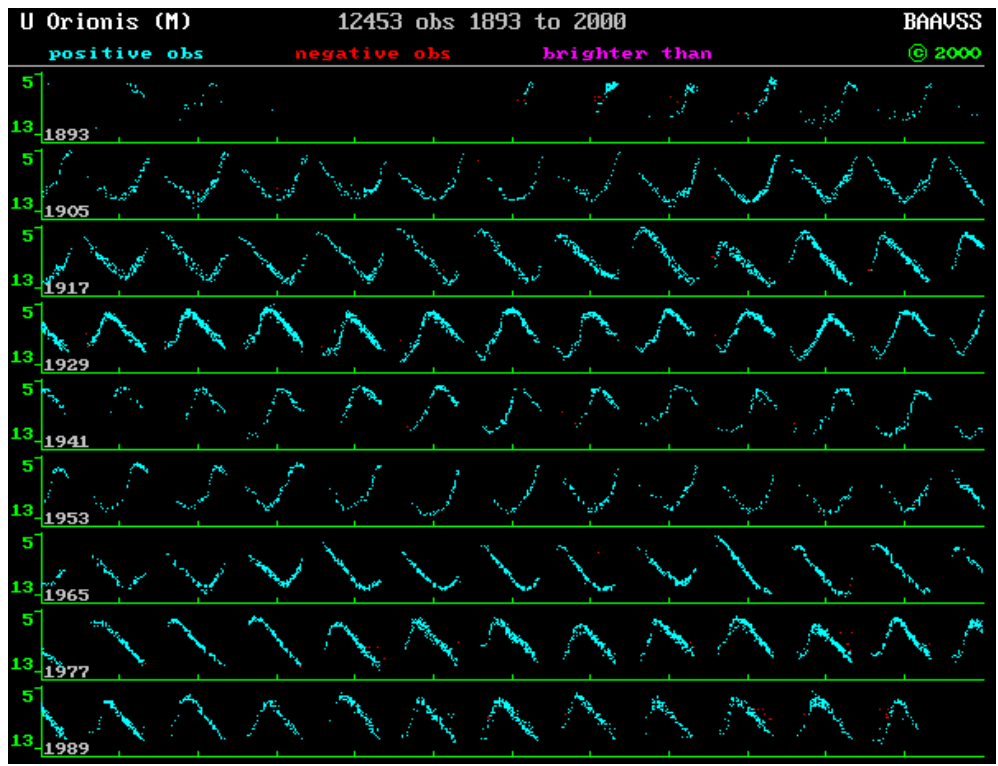

Figure 1: Light curve of the long period variable U Orionis between 1893 and 2000 for the original VSS database

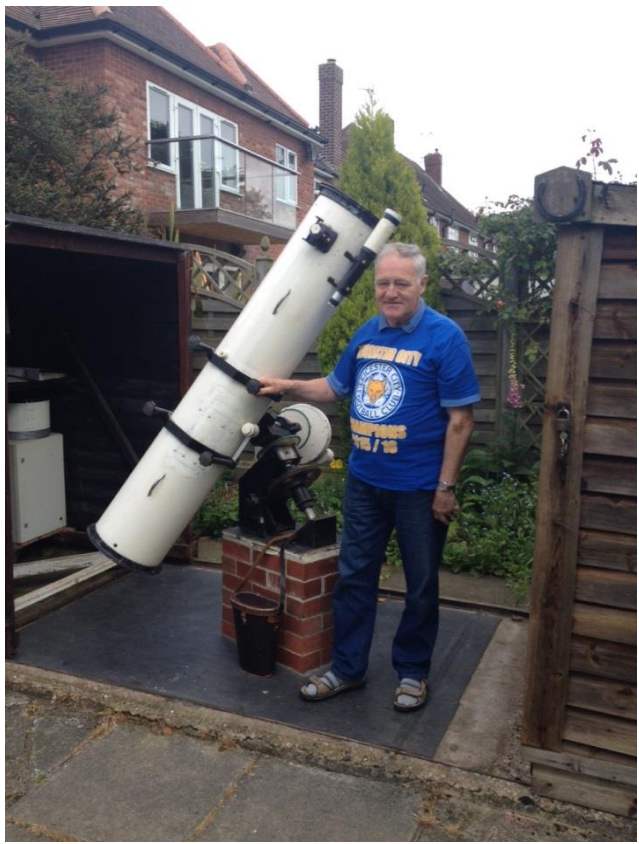

Figure 2: Mike Gainsford with his 25 cm Newtonian with which he made the one millionth observation in the VSS database on 1996 December 23

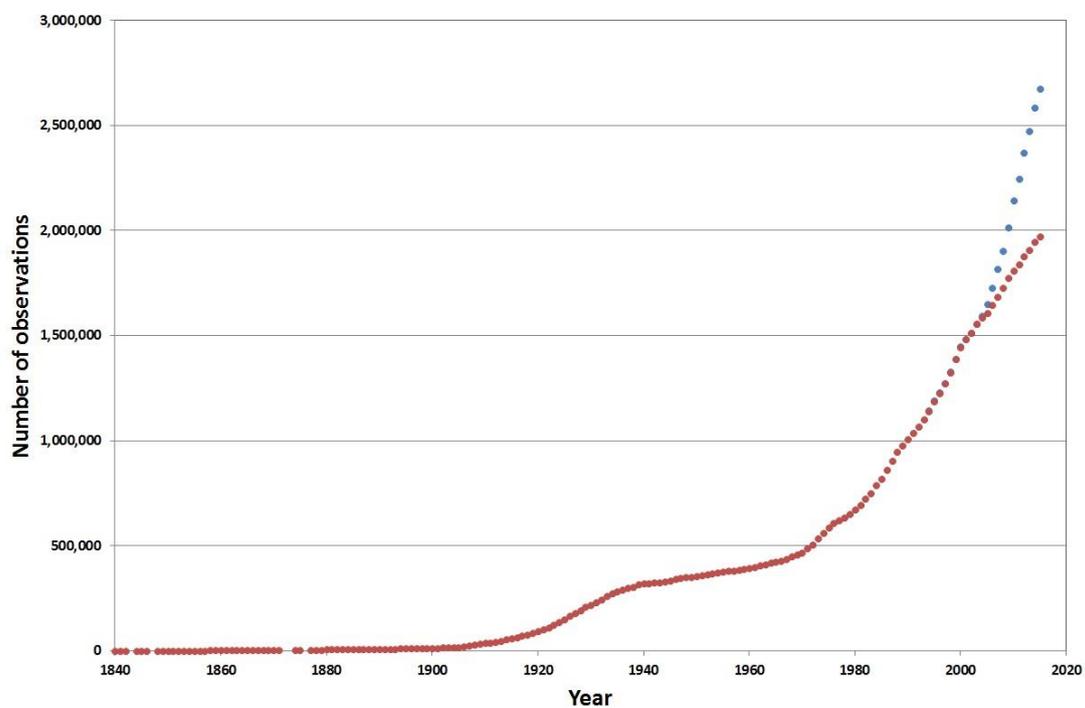

Figure 3: Cumulative graph of observations in the VSS database from 1840 to 2015. Red dots represent visual observations and blue dots are CCD observations

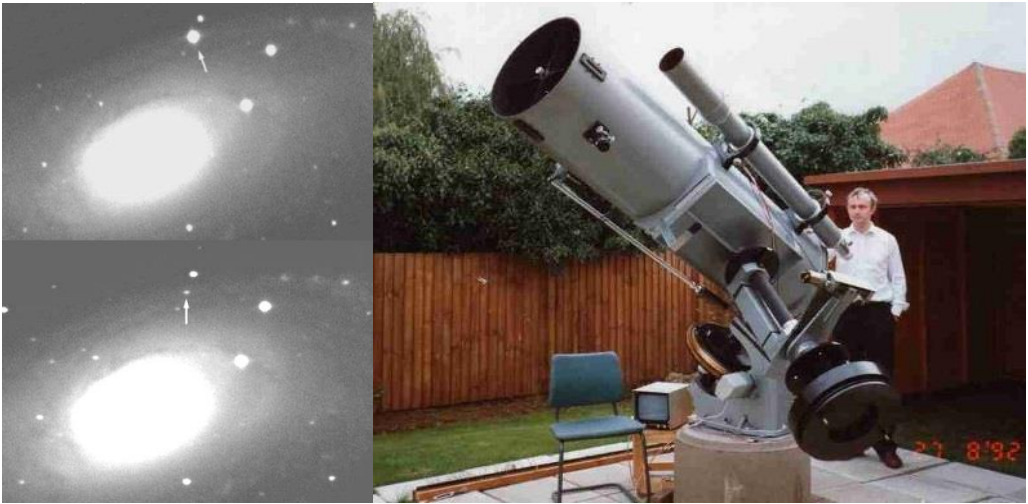

Figure 4 The dawn of CCD photometry (a) – left- Image of SN 1993J on 1993 April 15 (top) and 1994 Jan 17. (b) – right- Martin Mobberley and the 49 cm Newtonian used to take the CCD images

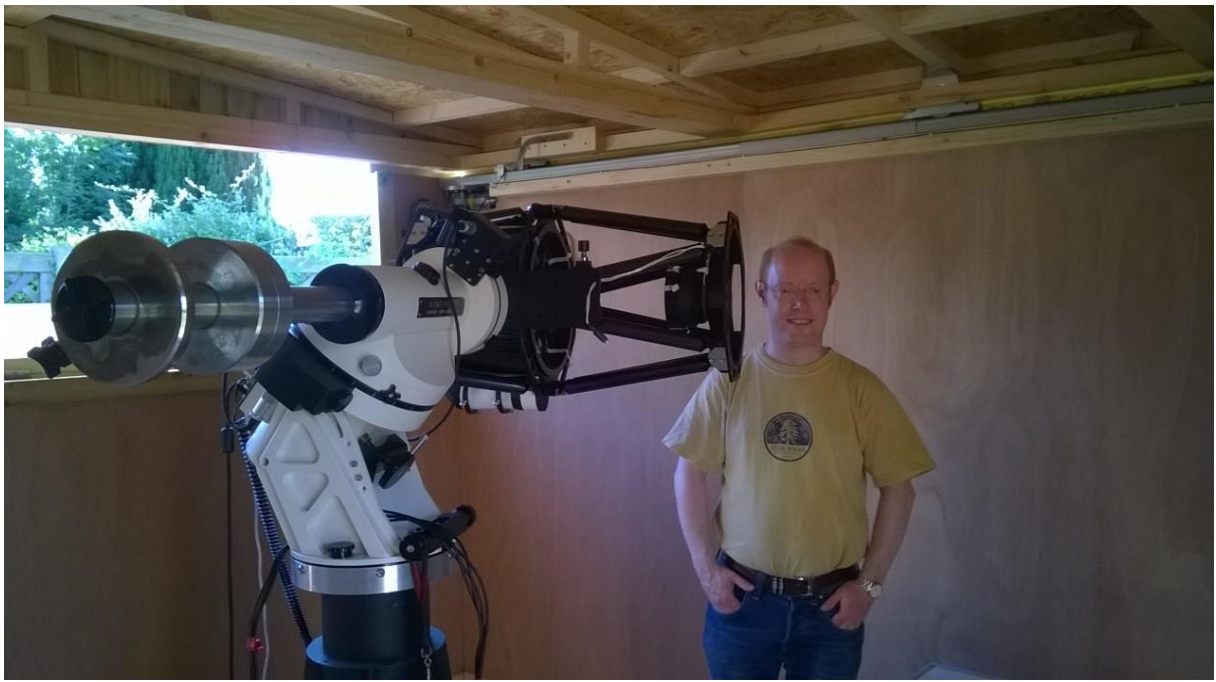

Figure 5: Andrew Wilson winner of the Association's Merlin medal for 2015

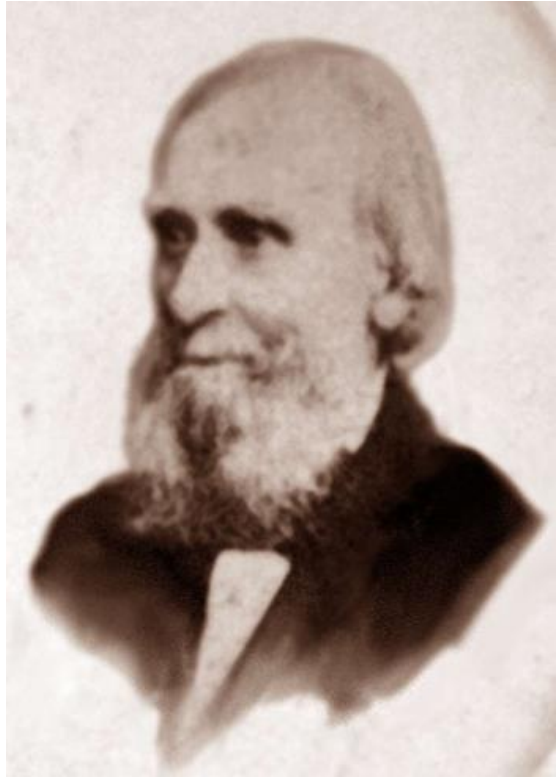

Figure 6: Joseph Baxendell (1815-1887)

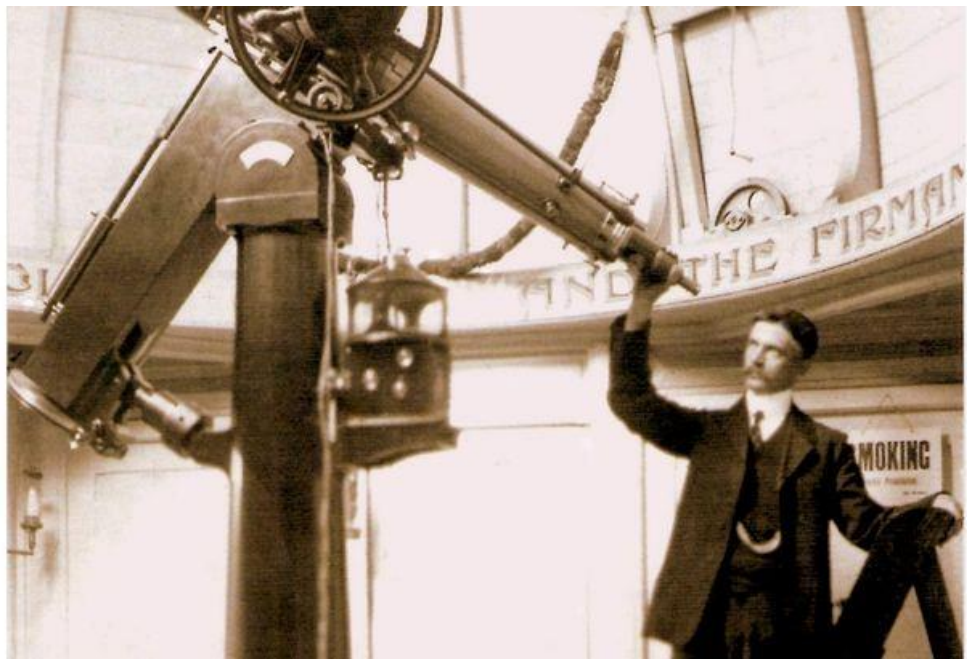

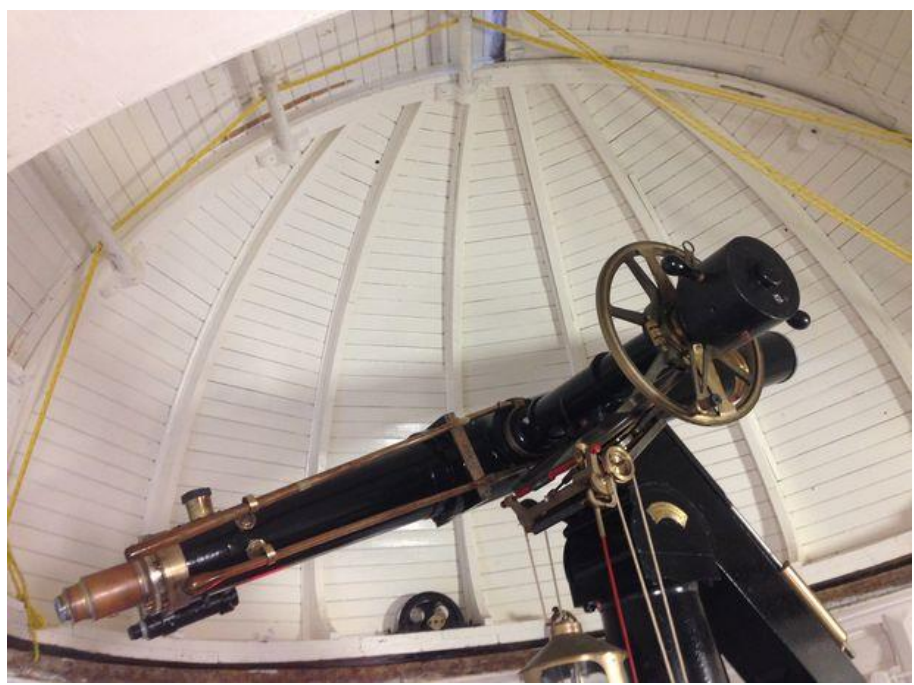

(a)

Figure 7: Baxendell's 6-inch (15 cm) Cooke refractor. (a) – top- with Joseph Baxendell Jr. ca. 1901. (b)- bottom- the instrument today

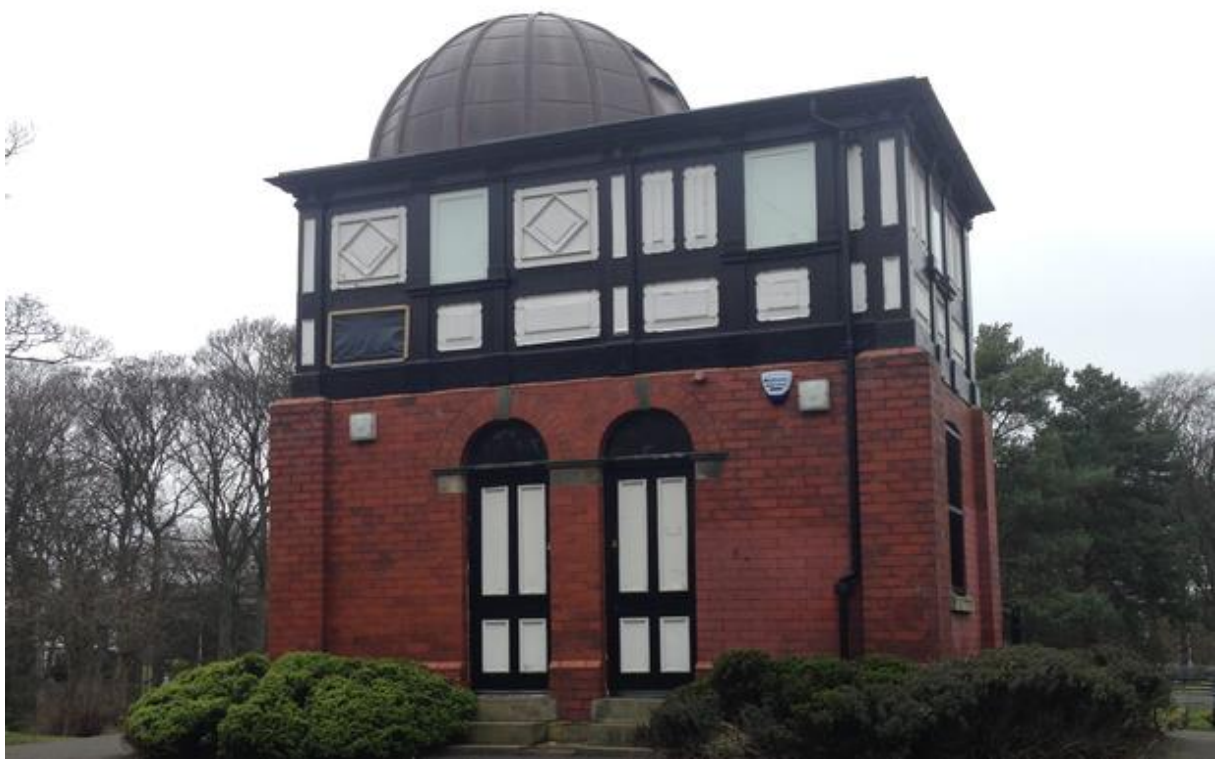

Figure 8: The Fernley Observatory in Hesketh Park, Southport

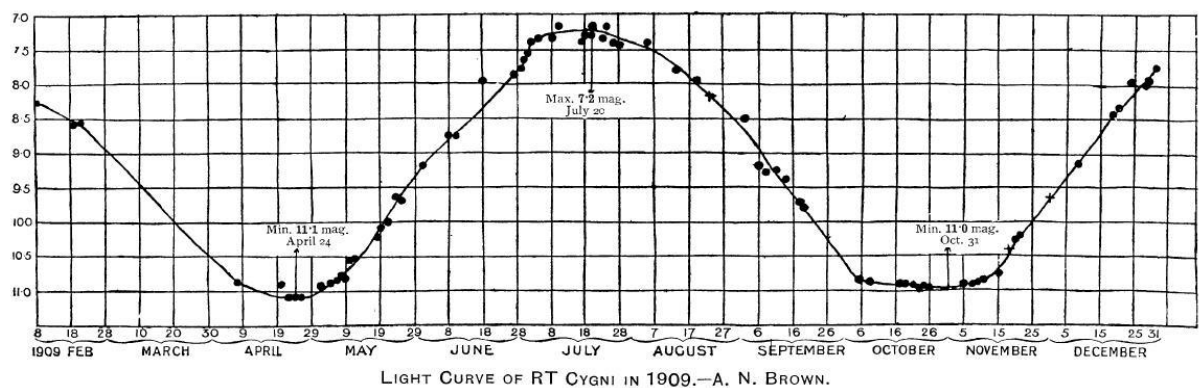

Figure 9: Hand drawn light curve of the LPV RT Cyg during by A.N. Brown, annotated with times of maximum and minimum

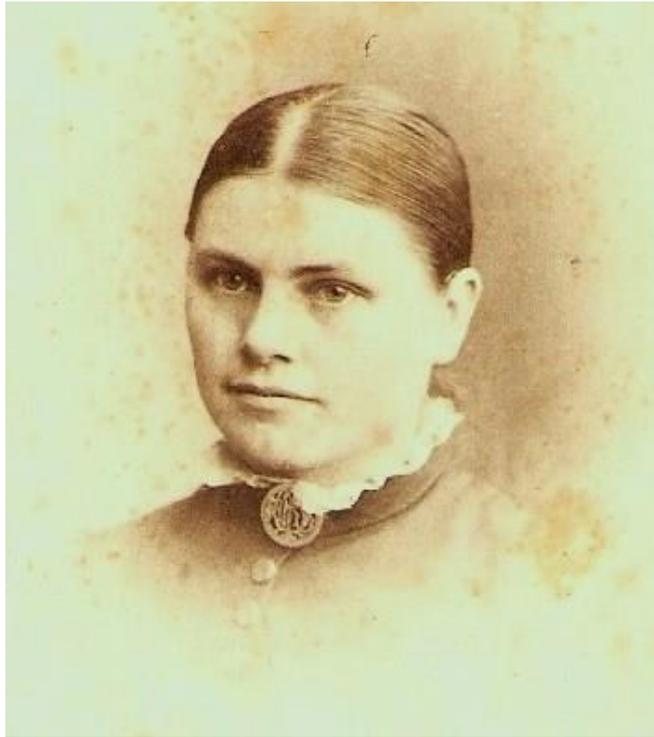

Figure 10: Mary Adela Blagg, FRAS (1858-1944)

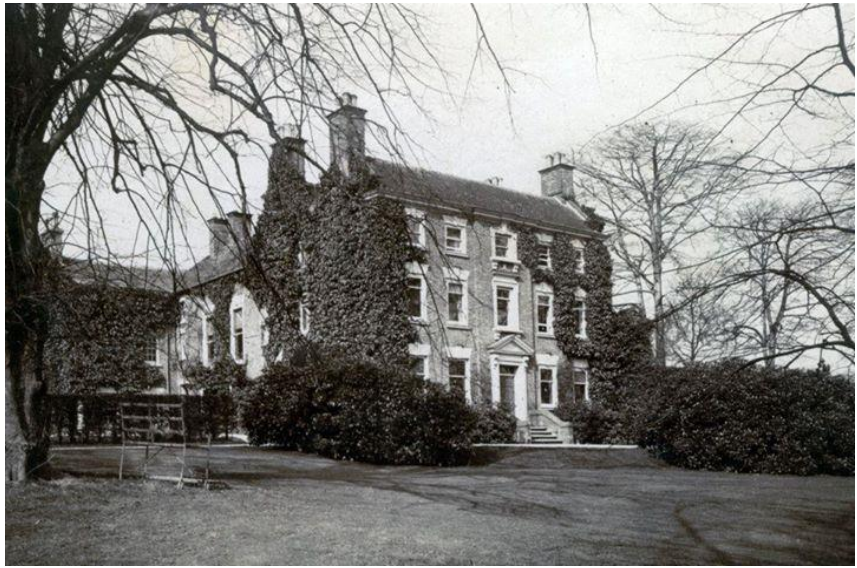

Figure 11: The Blagg family residence, Greenhill, Cheadle, Staffs (image courtesy of Richard Masefield)

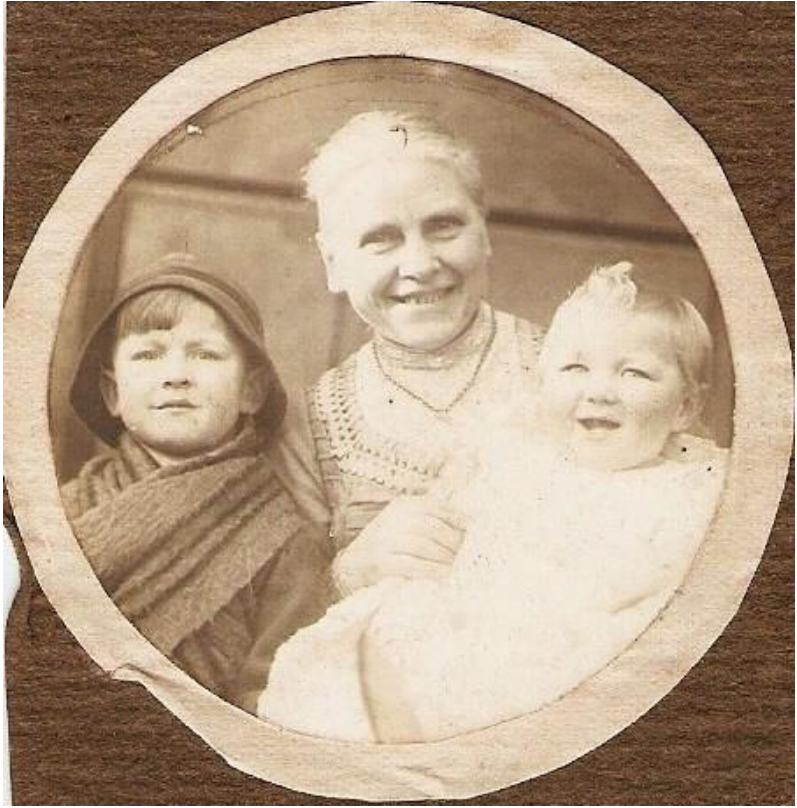

Figure 12: Mary Blagg pictured with 2 refugee children from Belgium, 1915 January 2 (image courtesy of Richard Masefield)

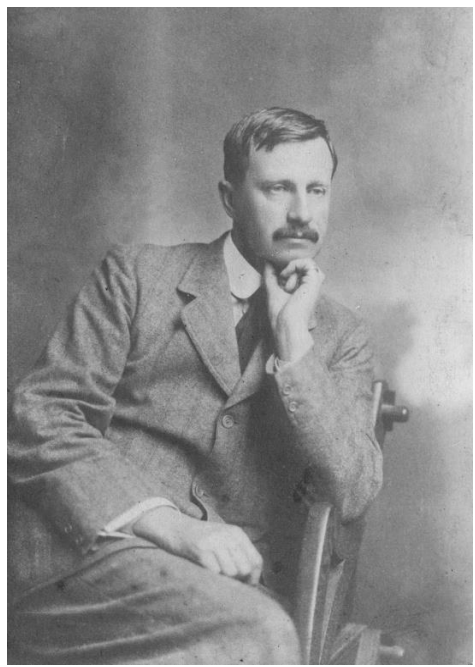

Figure 13: J.A. Hardcastle (1868-1917)

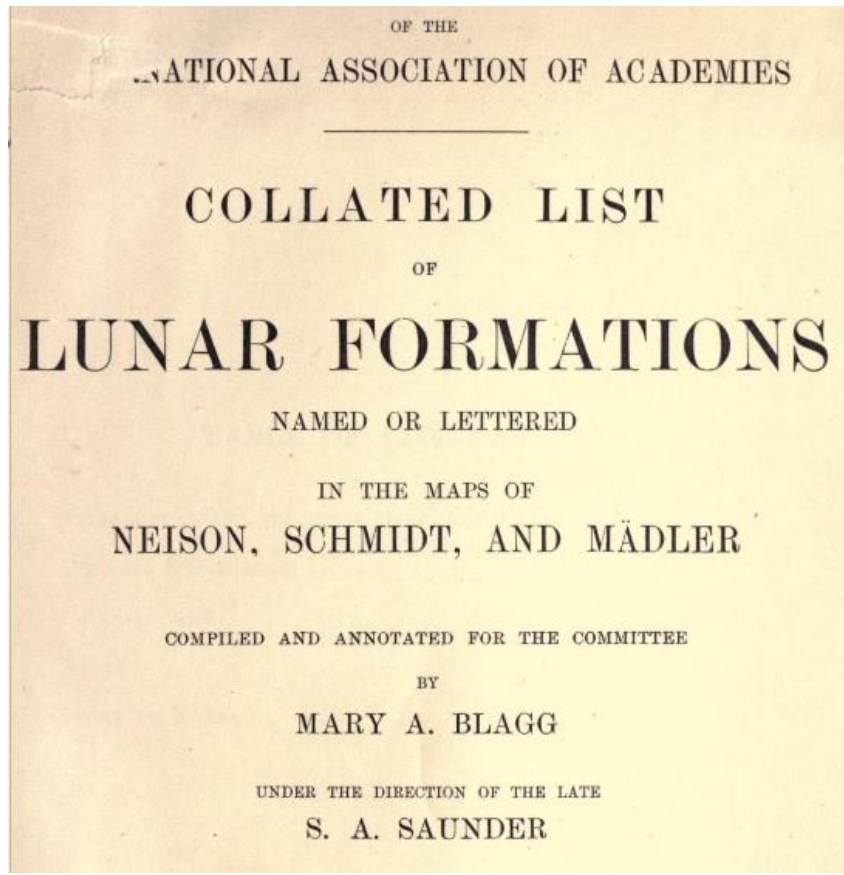

Figure 14: Title page of Blagg and Saunder's Collated List of Lunar formations (1913)

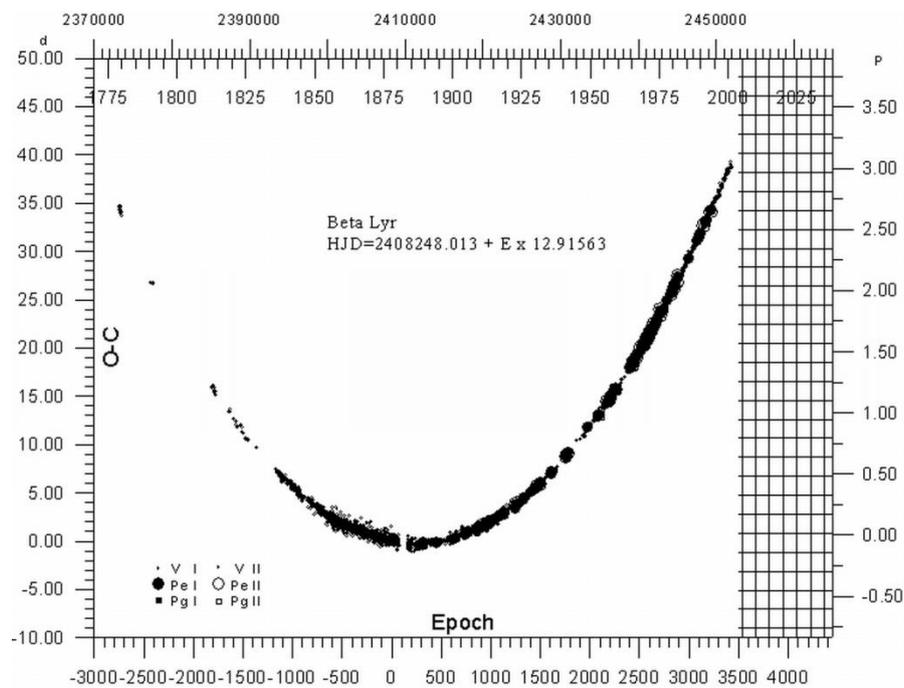

Figure 15: O-C diagram for β Lyr shows that the orbital period is increasing at a rate of about 19 seconds per year from reference (24)

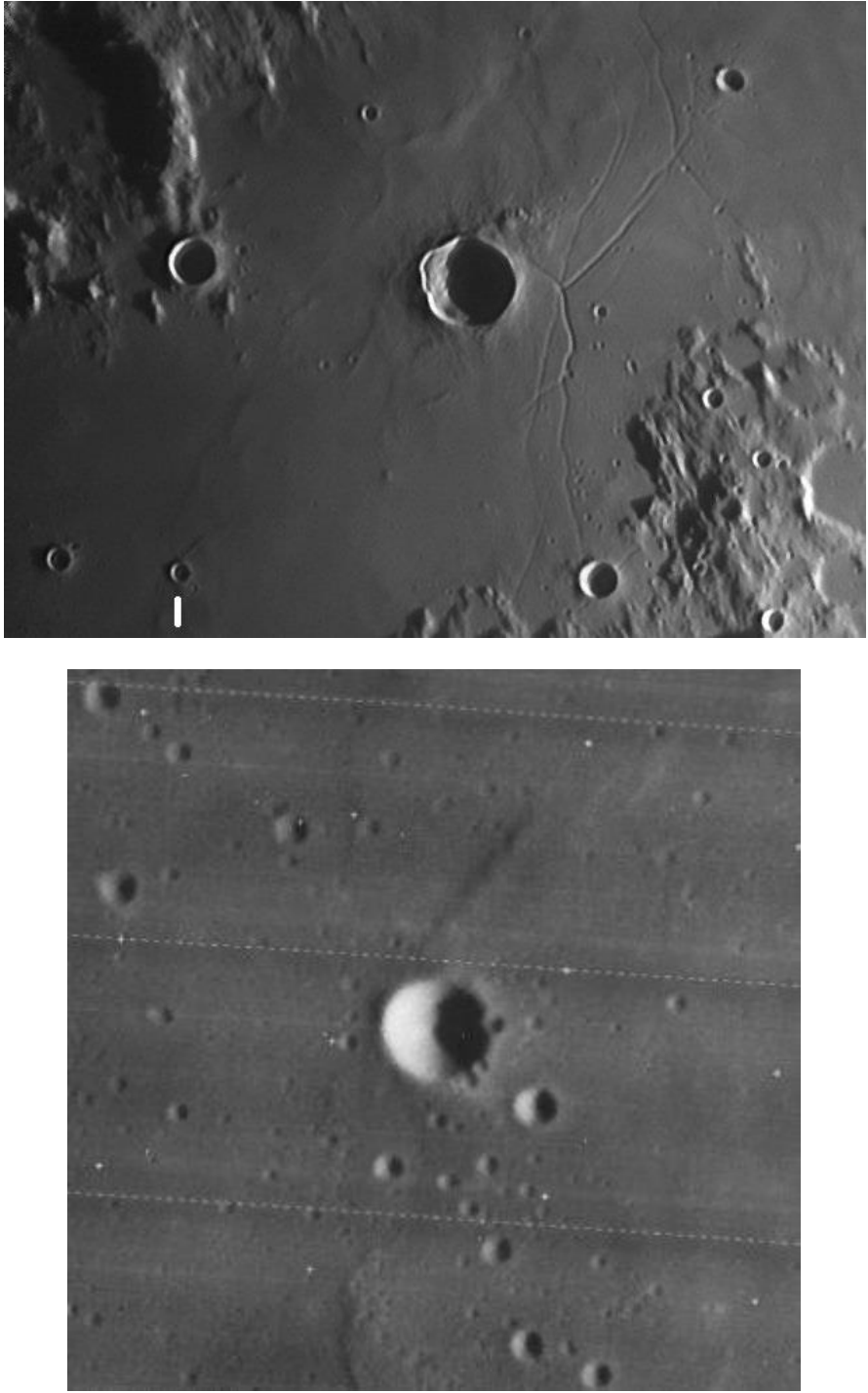

Figure 16: the lunar crater Blagg. (a) –top- The Triesnecker rille system, showing crater Blagg marked. Image by Professor Bill Leatherbarrow on 2017 January 5. 30 cm Maksutov-Cassegrain f/10 telescope with IR filter and ASI 290MM mono camera. (b) LRO image

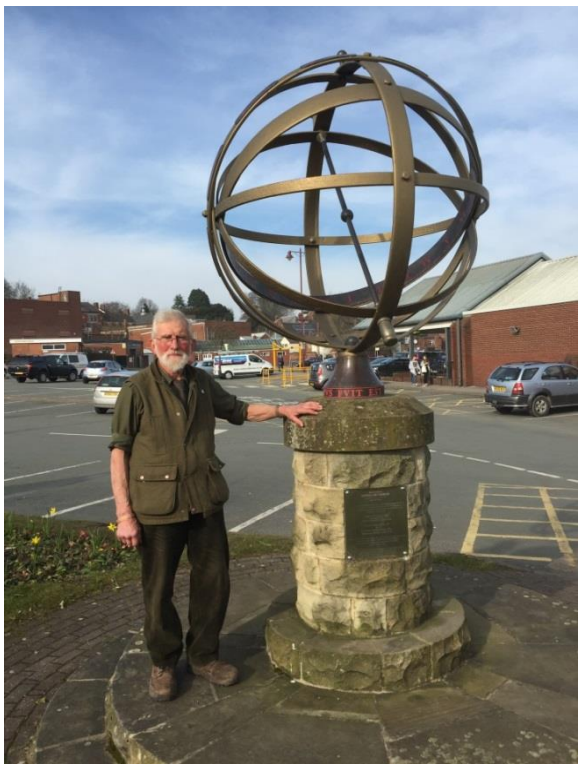

Figure 17: Armillary sphere at Cheadle, Staffordshire, constructed by James Plant (pictured) in memory of Mary Blagg

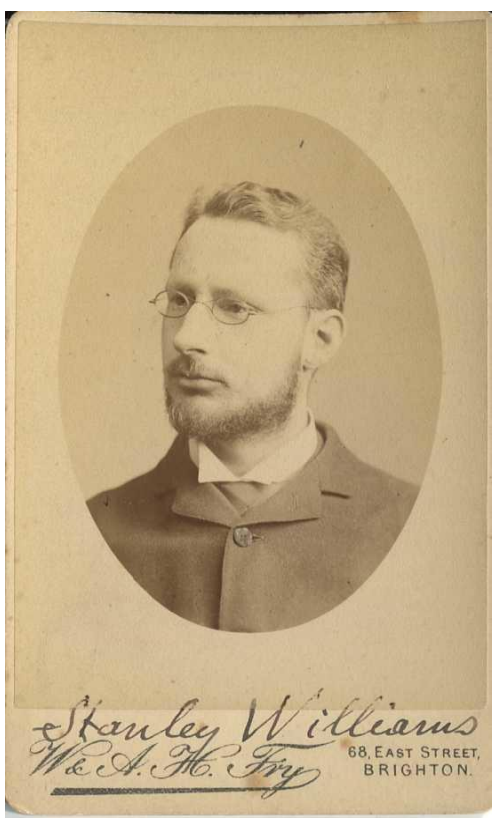

Figure 18: Signed photograph of Arthur Stanley Williams (1861-1938) in 1895 (43)

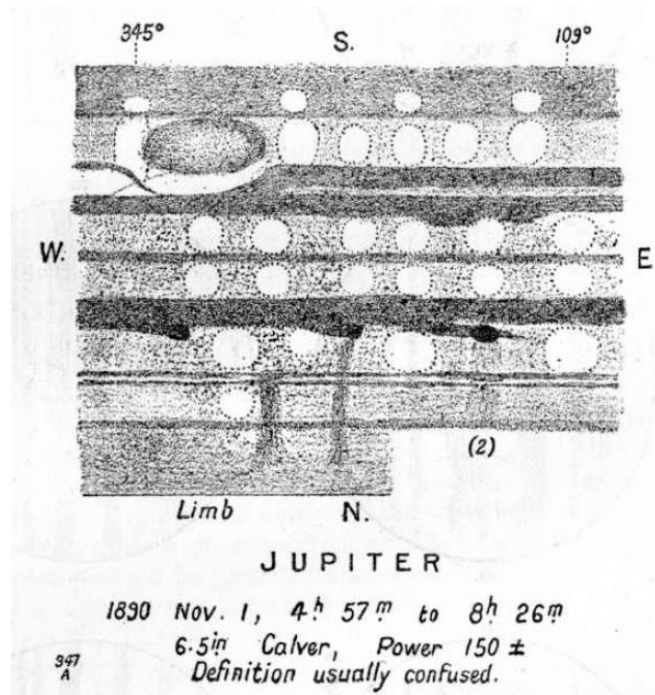

Figure 19 – Drawing of Jupiter by A. Stanley Williams, 1890 Nov 1 (44). This was made with the same 6½-inch (17 cm) Calver reflector as he used for his variable star observations.

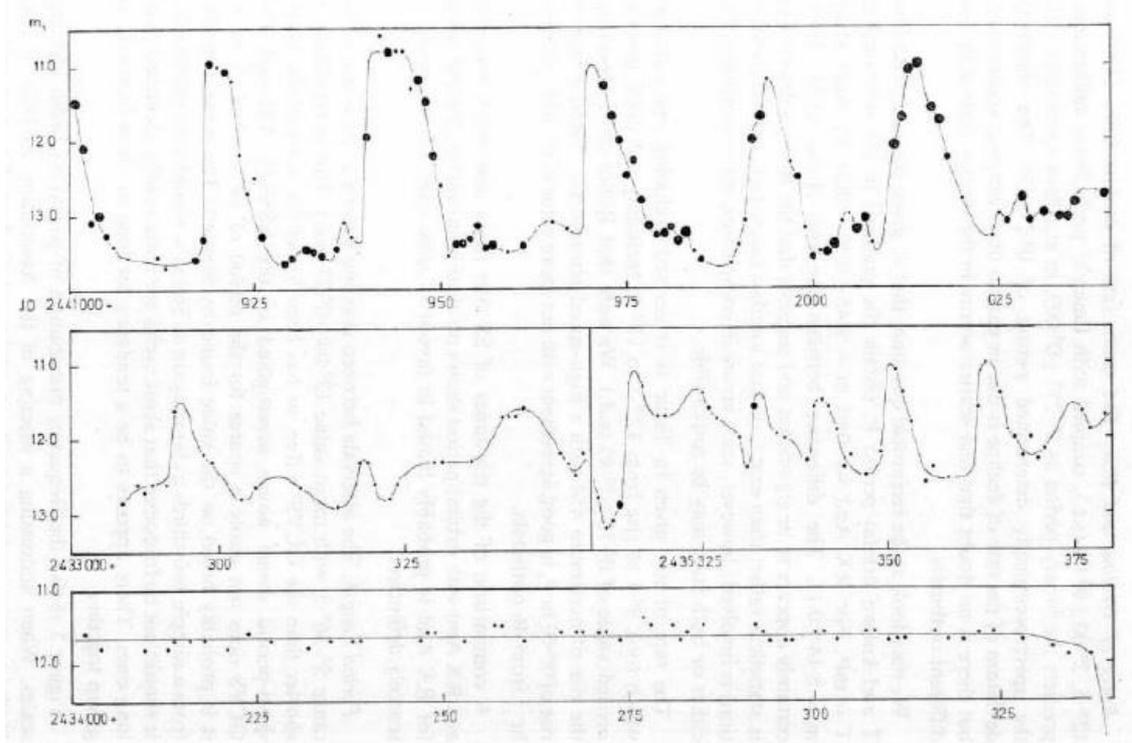

Figure 20: Light curve of RX And. Top: normal pattern of outburst behaviour in 1971. Middle: erratic behaviour in 1949 and 1955. Bottom: standstill last more than 130 days in 1952. Based on VSS data, from reference: (45)

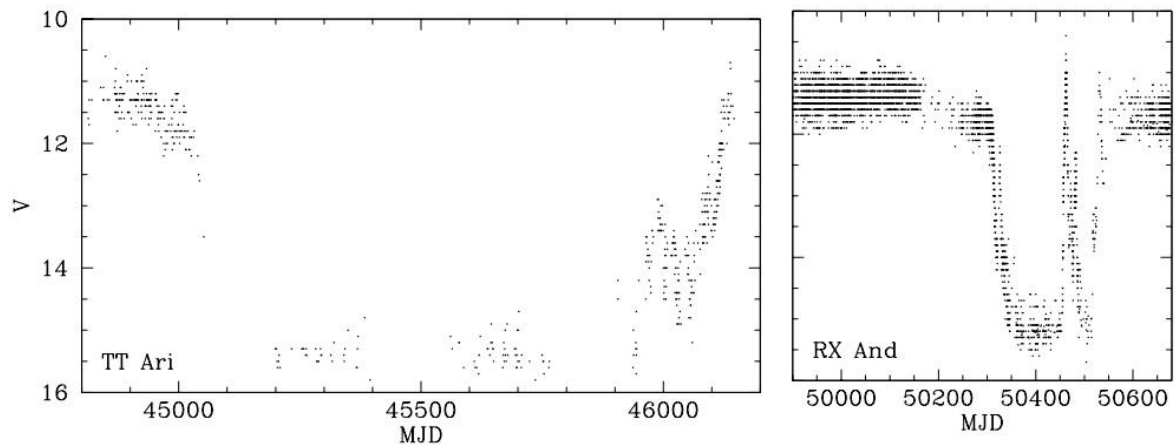

Figure 21: Light curves of TT Ari, a VY Scl star, and RX And each showing high and low states. From reference (30)

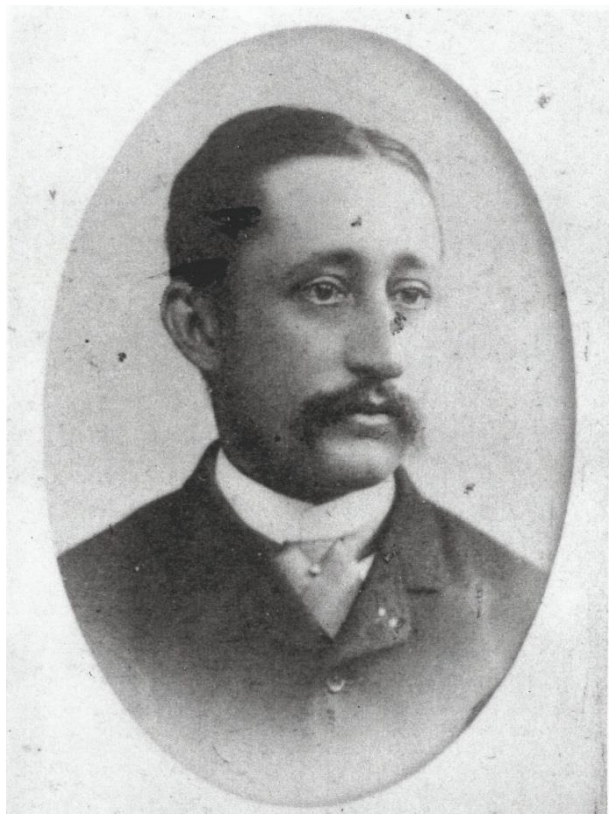

Figure 22: John Ellard Gore (1845-1910), first Director of the VSS

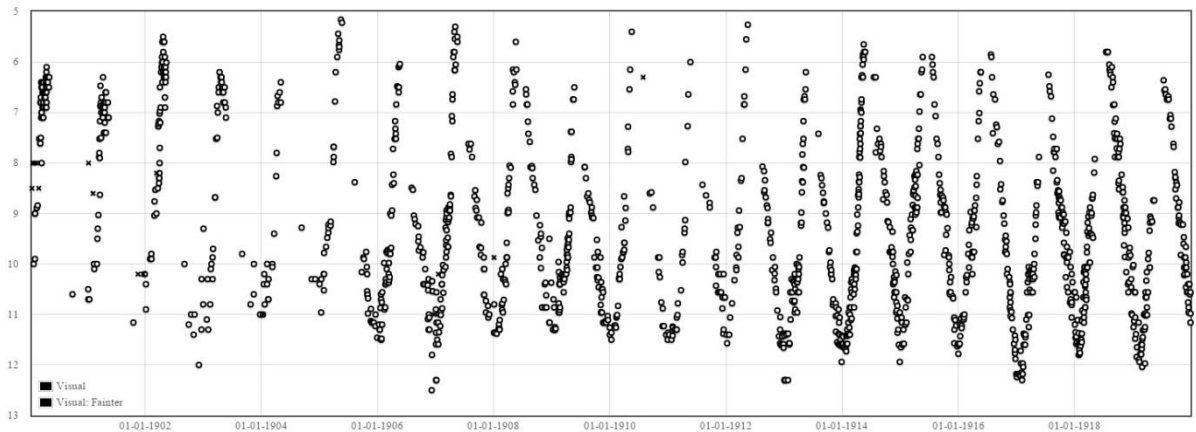

Figure 23: Light curve of the LPV U Ori between 1900 and 1919. The y-axis is the visual magnitude. See reference (46) for list of observers (VSS database)

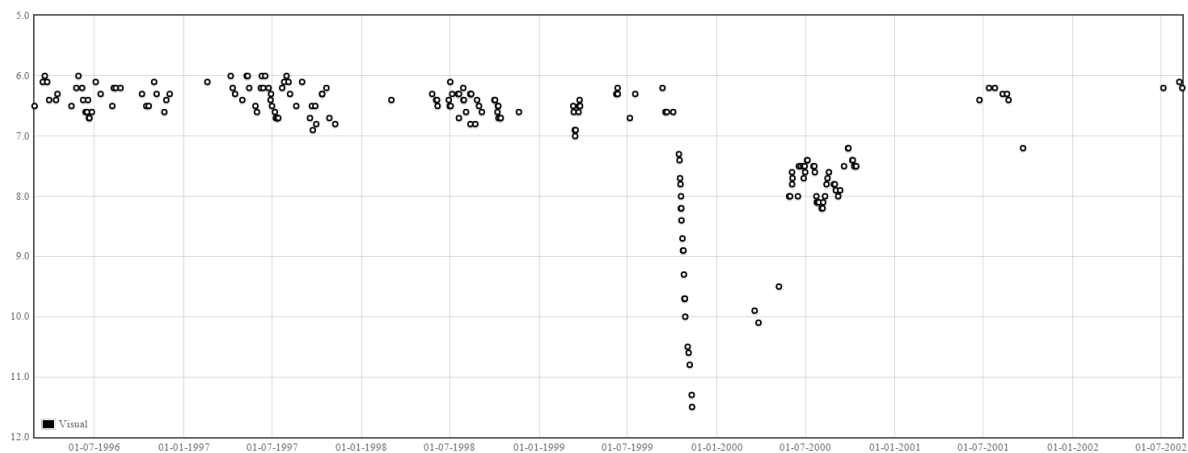

Figure 24: Visual observations of RY Sgr, 1996 to 2002 showing a deep fade in 1999. The y-axis is the visual magnitude. Observers: B. H. Granslo, C. Henshaw, G. Stefanopoulos, J. Toone, L. A. G. Monard, S. Otero (VSS database)

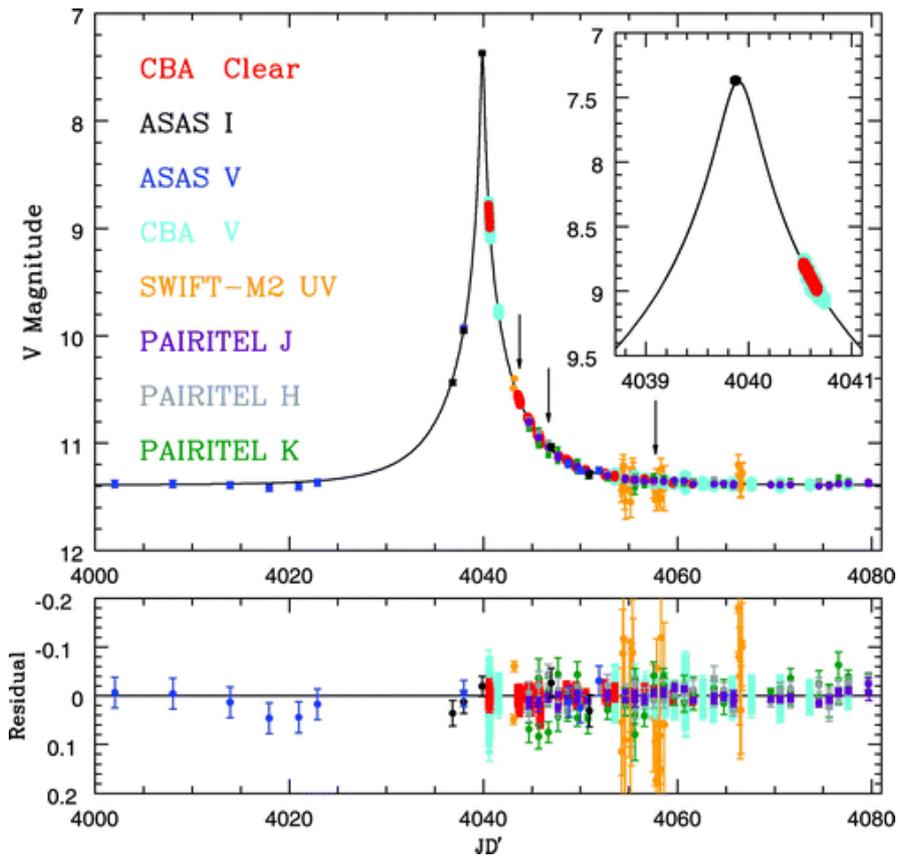

Figure 25: Light curve of GSC 3656-1328 during its brightening in 2006. The graph at the top shows Clear and V-band magnitudes (inset: detail near the peak of the event). The solid line shows the best-fit microlensing model. The plot at the bottom shows the residuals from this model. From reference (41)